\documentclass[manuscript]{acmart}

\usepackage{graphicx} 
\usepackage{tcolorbox}
\usepackage{xcolor,pifont}

\newcommand*\colourcheck[1]{%
  \expandafter\newcommand\csname #1check\endcsname{\textcolor{#1}{\ding{52}}}%
}
\colourcheck{green}

\usepackage[hide]{inlinecomments}

\newcommand{\rf}[1]{\textcolor{olive}{\nb{Robert}{#1}}}
\newcommand{\todo}[1]{\textcolor{orange}{\nb{TODO}{#1}}}

\definecolor{blueish}{cmyk}{61, 47, 0, 40}

\newcommand{\implication}[2]
{
\begin{tcolorbox}[colback=blueish!5!white,colframe=blueish!90!black,title=#1]

\emph{#2}

\end{tcolorbox}
}

\usepackage{customtables}

\usepackage{cleveref}

\newcommand{\etal}{\emph{et al.}\xspace}

\usepackage{svg}

\newcommand{\ie}{{\emph{i.e.},}\xspace}

\begin{document}

\title{Exploring the Garden of Forking Paths in Empirical Software Engineering Research: A Multiverse Analysis}

\author{Nathan Cassee}
\email{nathancassee@uvic.ca}

\author{Robert Feldt}
\email{robert.feldt@chalmers.se}

\renewcommand{\shortauthors}{Cassee and Feldt}

\begin{abstract}

In empirical software engineering (SE) research, researchers have considerable freedom to decide how to process data, what operationalizations to use, and which statistical model to fit. Gelman and Loken refer to this freedom as leading to a ``garden of forking paths''. Although this freedom is often seen as an advantage, it also poses a threat to robustness and replicability: variations in analytical decisions, even when justifiable, can lead to divergent conclusions. 

\draft{3}{To better understand this risk, we conducted a so-called \textit{multiverse analysis} on a published empirical SE paper. The paper we picked is a Mining Software Repositories study, as MSR studies commonly use non-trivial statistical models to analyze post-hoc, observational data.}  In the study, we identified nine pivotal analytical decisions--each with at least one equally defensible alternative--and systematically reran all the 3,072 resulting analysis pipelines on the original dataset. Interestingly, only 6 of these universes (<0.2\%) reproduced the published results; the overwhelming majority produced qualitatively different, and sometimes even opposite, findings.

\draft{5}{This case study of a data analytical method commonly applied to empirical software engineering data reveals} how methodological choices can exert a more profound influence on outcomes than is often acknowledged. We therefore advocate that SE researchers complement standard reporting with robustness checks across plausible analysis variants or, at least, explicitly justify each analytical decision. \draft{5}{We propose a structured classification model to help classify and improve justification for methodological choices.} Secondly, we show how the multiverse analysis is a practical tool in the methodological arsenal of SE researchers, one that can help produce more reliable, reproducible science.

\end{abstract}

\maketitle


\section{Introduction}

Since the early 2010s, science has faced a crisis of confidence. The so-called replication crisis---sparked by a series of failed replications of prominent psychology findings---has exposed deep flaws in how scientific studies are designed, analyzed, and reported~\cite{Chambers:2017}. The problems run deeper than fraudulent practices or overt P-hacking. Even in well-intentioned studies, the sheer number of methodological decisions researchers can make, described by Gelman and Loken as the ''garden of forking paths'',\footnote{Inspired by the short story of Jorge Luis Borges.} can silently steer results in different directions~\cite{Gelman:2013}. The garden of forking paths represents the idea that scientific findings, like statistical significance, emerge not because of manipulative intent but due to researchers' vast, often unacknowledged flexibility in analyzing their data. A striking illustration of this was provided by Silberzahn et al.~\cite{Silberzahn:2018}, who showed that different teams given the same dataset and research question arrived at vastly different conclusions, simply because they made different--but all reasonable--choices in how to analyze the data. This highlights how the garden of forking paths might influence study outcomes and why it is vital to study the effect of methodological decisions on outcomes.

One method that can be used to study methodological sensitivity, which is still underutilized in many fields, is the systematic mapping and investigation of how analytical choices influence study outcomes. This form of analysis, sometimes referred to as a \textit{multiverse analysis} after Steegen et al.\cite{Steegen:2016}, enables researchers to explore a structured set of alternative analytical paths. In doing so, it exposes how robust or fragile a study's conclusions are to reasonable variations in methodology. While such approaches have begun to gain traction in disciplines such as psychology and epidemiology, to our knowledge, they have not yet been applied in software engineering.

This gap is surprising, especially given that the software engineering (SE) community has actively engaged with other facets of the replication crisis~\cite{shepperd2018role,mendez2019open,Cockburn:2020,ernst2023registered}. 
In particular, the Mining Software Repositories (MSR) community has recognized the methodological challenges inherently present in the data. Although MSR researchers work with rich and powerful data sources~\cite{Bird:2009,Kalliamvakou:2016}, uniquely suited to studying software engineering phenomena~\cite{Ferreirra:2022,Liu:2022,MurphyHill:2021}, they must make many decisions to mitigate the known perils associated with such data~\cite {Kalliamvakou:2016,Bird:2009,Munaiah:2017}.

Methodological diversity is common in software engineering literature. For instance, a meta-study of MSR studies by Mahadi et al.\cite{Mahadi:2020} found considerable variation in how studies addressing the same research question operationalized their analyses. Wyrich et al.\cite{Wyrich:2024} demonstrated how even slight differences in how SE researchers define key constructs can lead to results that are difficult to compare and potentially contradictory, and \draft{3}{re-analysis of previously reported findings using different analytical methods has already shown how findings can change~\cite{Frattini:2024,Furia:2023}}.

In this work, we want to further understand the impact of these methodological variations, and therefore, we pose:

    \begin{rqs}{}
        RQ & How sensitive are the conclusions of Empirical Software Engineering studies to methodological decisions? \\
        
    \end{rqs}

To answer this question, we selected a published MSR study~\cite{Cassee:2020} that uses a data-analytical method employed in over ten MSR studies and known for affording researchers a high degree of freedom, making it a fitting choice for our investigation. 
We systematically explore 3,072 distinct, yet plausible, analytical variants of the original study--each representing different combinations of nine methodological decisions. 
Our goal is to quantify how often and to what extent the study's conclusions change when alternative, yet defensible, analytical choices are made. 

The results are instructive, even in this single case.
Among the 3,072 analytic paths we explored, only six (0.2\%) reproduced the original study’s result. Many paths yielded null or even contradictory outcomes. Each of the nine analytical decisions had the power to flip the result---underscoring the fragility of conclusions drawn. \draft{3}{Our findings serve as a cautionary tale, \textit{as we find that if researchers' degrees of freedom increase, confidence in results decreases}. Especially because empirical software engineering, and MSR in particular, rely on data sources that require a high degree of researcher freedom.}

\draft{2}{More constructively, our study highlights the need for greater transparency in \textbf{justifying} methodological decisions in MSR research. We introduce a model that can be used to reason about the different types of justification and discuss practices to strengthen them.
Moreover, our findings demonstrate the value of multiverse analyses: by systematically exploring alternative methodological choices--particularly when researcher degrees of freedom are high--one can pinpoint which methodological decisions, if any, most critically affect results.
}

\section{Related Work}

\draft{2}{The factors contributing to the replication crisis have been studied extensively. \citet{Chambers:2017} identified seven ``sins'' that capture problematic research practices. A range of potential solutions has also been proposed~\cite{ernst2023registered,Trafimow:2018}. \draft{5}{In this section, we focus on prior work related to methodological freedom, statistical analysis in software engineering, and multiverse analysis.}}

\draft{5}{The validity and reliability of empirical software engineering literature has been studied extensively. Early work by \citet{Dyba:2006} in 2006 already showed how the sample size in existing software engineering experiments was too low. Similarly, \citet{Reyes:2018} shows that many experimental software engineering papers make statistical errors seen in other disciplines. Understanding the statistical analysis reported in software engineering is further complicated by inconsistent reporting guidelines. Both \citet{Neto:2019,Santos:2021} describe how heterogeneity and inconsistent reporting guidelines of statistical tests complicate any sort of meta-analysis. To help remediate some of the issues reported previously, guidelines on how to apply statistical methods have been described~\cite{Arcuri:2014,Kitchenham:2017}. However, these guidelines often focus on how to report and visualize results rather than on methodological freedom.  }

Across several fields, researchers have shown how degrees of methodological freedom can lead to varying outcomes.  
\citet{Silberzahn:2018} report that 29 independent analysis teams, tasked with answering the same research question, employed a wide range of analytical methods and reached conflicting conclusions. 
Similarly, \citet{Schweinsberg:2021} demonstrates that when given substantial flexibility in data analysis, different teams make divergent choices, producing inconsistent results.
\citet{Sarstedt:2024} further show that even when teams analyze the same model, their different decisions about data processing lead to varying effect sizes.

To address the link between methodological freedom and study outcomes, \citet{Steegen:2016} introduced the concept of \emph{multiverse analysis}.
\citet{DelGiudice:2021} describes guidelines for conducting meaningful multiverse analyses, emphasizing the need to explore only reasonable methodological alternatives. \draft{3}{\citet{Harder:2020} discusses how multiverses should be expanded to not just include data analytical decisions, but also decisions related to data collection. While \citet{Simonsohn:2020} introduces specification curve analysis, a method that uses multiverses to make inferences about the underlying data.}
To help interpret multiverses, \citet{Dragicevic:2019} introduces a tool that can interactively explore how methodological choices affect outcomes. \draft{5}{Similarly, \citet{Boba:2021} presents a formalized DSL to systematically explore multiverses and assist in multiverse analysis. }
\citet{Bell:2022} apply the multiverse concept to experimental design choices in machine learning benchmarks, proposing a framework to strengthen benchmarking robustness. 

To our knowledge, no multiverse analyses have been conducted specifically in software engineering. \draft{5}{While sensitivity analyses~\cite{Saltelli:2019} can also appear similar in nature to multiverse analyses, there are several key differences. Where a sensitivity analysis is often used to show that a specific assumption does not bias the outcome, the point of a multiverse analysis is to show how the freedom of researchers to make analytical decisions influences outcomes.}

While no explicit multiverses have been conducted in software engineering, the field has discussed and hinted at the adverse effects of methodological freedom.
In a replication study, \citet{Mahadi:2020} observe that different studies addressing the same research question make numerous, varying methodological choices, complicating direct comparisons. 
Similarly, \citet{Shepperd:2014}, finds conflicting results across defect prediction studies, and \citet{Wessel2022Quality} found that different Regression Discontinuity Design studies report conflicting or incomparable results.

Thus, while prior software engineering research has documented heterogeneity in methodological decisions, it has not systematically examined how sensitive study outcomes are to those choices. This gap is the focus of the present work.

\section{Study Overview}
\label{sec:background}

To understand how analytical decisions influence the outcome of Mining Software Repositories studies, we conduct a \emph{multiverse} analysis~\cite{Steegen:2016}. In this multiverse analysis, we study alternatives (\emph{universes}) to analytical \emph{decisions} made in an MSR study. By recording the outcomes (significance scores) of the study, we learn whether there is any relation between data analytical decisions and study outcomes 

\begin{figure}
    \centering
    \includegraphics[width=0.95\linewidth]{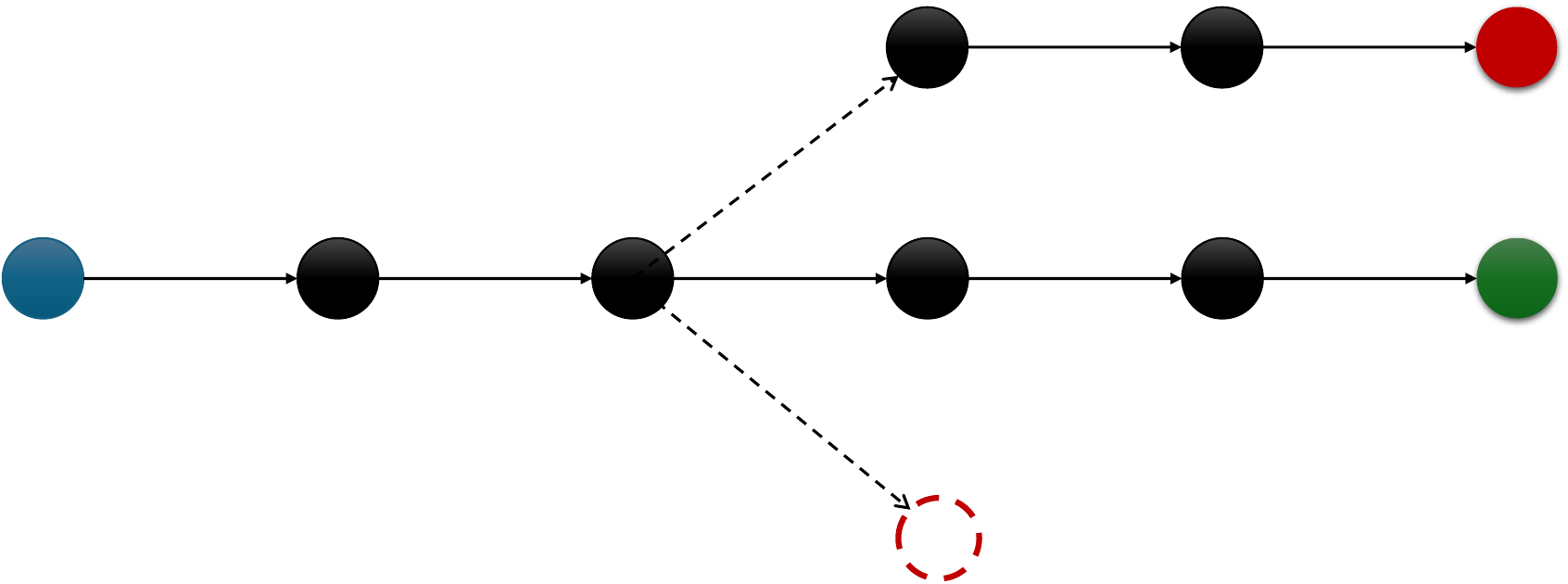}
    \caption{Visualization of the core idea of a multiverse analysis. A study moves from research question (blue node) to outcome (green node) through a path of methodological decisions (black nodes). In a multiverse analysis, alternative paths (dashed arrows) are explored to study whether these alternatives change outcomes (red node).   }
    \label{fig:decision-graph}
\end{figure}

\Cref{fig:decision-graph} visualizes the core idea of a multiverse analysis. \draft{5}{A multiverse analysis explores whether alternative methodological decisions can result in alternative outcomes. \textbf{In a multiverse analysis, a single \emph{universe} represents one set of analytical decisions leading from research question to outcome. Meanwhile, the multiverse is a set of universes representing the valid methodological designs to address a research question. }
Through a systematic exploration of these universes, created by identifying alternative choices for methodological decisions, we re-examine the original research question.} The goal of such an analysis is to identify the sensitivity of the outcome to methodological decisions.   

\draft{5}{Many studies in empirical software engineering consist of a large number of methodological decisions, with many alternatives to these decisions that are typically considered~\cite{Robillard:2024}. Which is why we believe it is important to apply multiverse analyses. However, it's important to note that not every alternative is reasonable~\cite{DelGiudice:2021}.} For instance, deciding to use a parametric test on non-parametric data is an example of an alternative we are not interested in exploring -- as the alternative (parametric test) does not meet existing assumptions and is \draft{5}{known to produce potentially invalid outcomes. Therefore, we carefully construct the universes we explore in this multiverse analysis.} 

However, before starting the analysis, we first pick and describe a data analytical method commonly applied to MSR data. Then we pick a primary study that applies this method to empirical software engineering data. In the remainder of this section, we first provide background information on the data analytical method \Cref{sec:background-rdit}; we discuss how this method has been applied to software engineering and how there is quite a lot of heterogeneity in the decisions made when this method is used to study software engineering (\Cref{sec:rw-rdd}). Finally, we pick a single case study that applies this data analytical method, and we describe it (\Cref{sec:primary-study}).

\subsection{Regression Discontinuity in Time}
\label{sec:background-rdit}

Studying how an intervention impacts a process is a challenging problem. Regression Discontinuity in Time (RDiT)~\cite{Hausman:2018} is one of the statistical methods applied to observational data to study the impact of interventions.
RDiT is a quasi-experimental statistical technique applied to observational data in settings where it is impractical or impossible to conduct randomized trials. A commonly used example motivating the use of RDiT is a situation like a power plant installing pollution filters, and scientists wishing to understand whether this reduces pollution in the surrounding plant environment~\cite{Hausman:2018}. These are settings where studying the same power plant simultaneously with and without the filter is impossible. Furthermore, as there are usually many sources of pollution, there is insufficient control over the environment to isolate the effect of the power plant on the environment. In those cases, a quasi-experimental technique like RDiT can help quantify the impact of the pollution filters.
In the literature, this method is also known as regression discontinuity design (RDD); however, in this manuscript, we will refer to it as RDiT.

\begin{figure}[t]
    \centering
    \includegraphics[width=.9\linewidth]{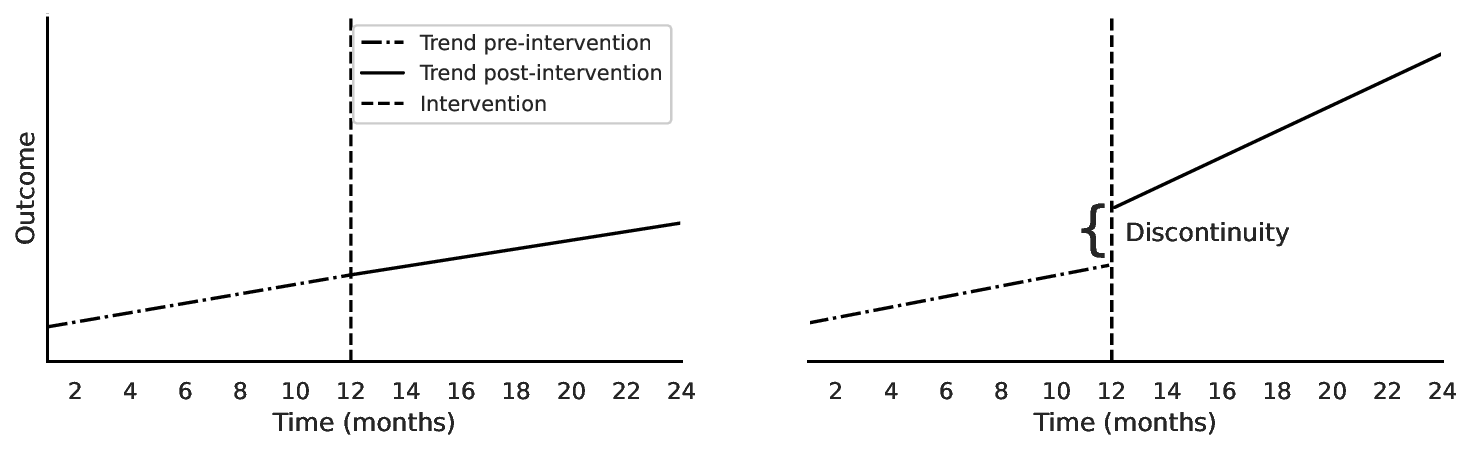}
    \caption{An example of Regression Discontinuity in Time (RDiT) design. \draft{5}{The plot shows two scenarios; in both scenarios, there is an intervention taking place in the 12th month. The left plot shows a scenario where the intervention does not have any effect, with no change in the outcome after the intervention, while the right plot illustrates a scenario where the intervention results in an observable change, with a visible discontinuity both in slope and intercept of the outcome.  }}
    \label{fig:rdit-example}
\end{figure}

In an RDiT design, two separate regressions are fit to a timeline. One regression fits the pre-intervention data points, while the second regression fits the post-intervention data points. The difference between slopes and intercepts of the two regressions is used to approximate the effect of an intervention. 
By comparing the two regressions, one can understand whether there was an immediate ``discontinuity'' post-intervention and whether there was a change in the trend.
\Cref{fig:rdit-example} shows a visual example of two RDiT models. In both plots, the x-axis represents time, which is usual in RDiT designs. The timeframes are centered around the intervention point, and in each plot, the two lines show the two fitted regression models.
The difference between the two plots is that the right plot suggests a "discontinuity" immediately after the intervention and a trend change post-intervention. Based on the left plot, one would conclude that the intervention had no effect, whereas, for the right plot, one would conclude that the intervention led to an increase in the dependent variable.  

\subsection{Regression Discontinuity in Time in Software Engineering}
\label{sec:rw-rdd}

Because of the challenges in studying interventions in software engineering, RDiT has been used to study the effect of interventions on various software engineering activities. In this section, we give an overview of these studies, and we highlight how there is a large variation in the data analytical decisions made in each of these studies. \draft{5}{To find RDiT studies in software engineering, we used an informal search process, combining forwards snowballing from two of the first reported RDiT studies in software engineering (\cite{Zhao:2017,Cassee:2020}) combined with searches on Google Scholar using keywords like \emph{``Regression Discontinuity Design''} and \emph{``Regression Discontinuity in Time''}. }

\begin{fancylongtblr}
    [
        caption = {The time modeling choices made in RDiT studies.},
        label = {tab:time-modeling-choices}
    ]
    {
    colspec = {X[2]X[3]X[1,r] l X[1,r] l X[2] l},
    }

    Authors & Topic & \# Periods & & Period Length & & Exclusion & \\
    Zhao \etal \cite{Zhao:2017} & Impact of CI on Commits \& PRs & 24 & & 30 days & & Middle period excluded & \greencheck \\
    Cassee \etal \cite{Cassee:2020} & Impact of CI on Code Reviews & 24 & \greencheck & 30 days &  & Middle period excluded & \greencheck \\ 
    Guo and Leitner \cite{Guo:2019} & Impact of CI on merge time & Variable & & 7 days & & No exclusion & \\ 
    Kavalar \etal \cite{Kavalar:2019} & Impact of QA tools on issues, churn, PRs \& contributors & Variable &  & 30 days & & One Month & \\
    Wessel \etal \cite{Wessel:2020} & Impact of bots on software engineering & 24 & & 30 days & & Middle period excluded & \\ 
    Kinsman \etal \cite{Kinsman:2021} & Workflows & 12 & & 30 days & & Middle period excluded & \greencheck \\ 
    Trockman \etal \cite{Trockman:2018} & Impact of badges on dependency age & 18 & & 30 days & & \emph{No Exclusion} & \\
    Zimmermann \etal \cite{Zimmermann:2019} & Impact of switching on bug trackers & 175 and 511 (days) & \greencheck & 1 day or 1 week & & \emph{No Exclusion} & \\ 
    Moldon \etal \cite{Moldon:2021} & Removal of GH Features & 2, 4, 6 & & weeks & & \emph{No Exclusion}. & \\ 
    Walden \etal \cite{Walden:2020} & Impact of security bugs on development & 50 & \greencheck & 30 days & & \emph{No Exclusion} & \\ 
    Moharil \etal \cite{Moharil:2022} & Impact of bot on issues & 24 & & 30 days & & Middle excluded & \\ 
    Saraiva \etal \cite{Saraiva:2023} & Impact of CI on code coverage & 24 & \greencheck & 30 days & & No exclusion & \greencheck \\ 
    Li \etal \cite{Li:2023} & Impact of issue report templates & 24 & & 30 days & & \emph{No Exclusion} & \\
    Ayoup \etal \cite{Ayoup:2022} & Impact of Gamyfing features on DevOps & 24 & & 30 days & & \emph{No Exclusion} & \\

\end{fancylongtblr}

\Cref{tab:time-modeling-choices} lists RDiT studies conducted in software engineering and the time modeling choices made in each of these studies. 
When a paper did not mention details, \emph{emphasis} is used to denote that no mention of any exclusion period was made. Some papers modeled multiple time series or used different decisions to fit several models. \draft{5}{The green checkmarks to the right of a chosen value indicate that the paper \textbf{includes} an explicit motivation for choosing that value. In the package of additional materials, we've included an Excel sheet with verbatim copies of the motivations from the listed studies. }

\draft{2}{To conduct an RDiT study, researchers need to make several data analytical decisions that determine how time is modeled.}
As seen in \Cref{tab:time-modeling-choices} there is a large variation across these studies in modeling time and excluding time periods to account for instability. Moreover, in each referenced study, justification for the choice of the period studied is often absent or brief. While it is acknowledged that the studied period shouldn't be too long, as mentioned in RDiT literature \cite{Hausman:2018}, or that some datapoints should be excluded to account for instability surrounding the intervention, the choice of popular options (excluding one-time period, analyzing 24 periods, and using 30 day long periods) is seldom justified, or only justified by referring to previously published studies. 

This heterogeneity is one of the reasons why we focus on RDiT: \draft{3}{To apply RDiT, a researcher needs to make many methodological decisions}. \Cref{tab:time-modeling-choices} shows \draft{3}{researchers actively use freedom to make} different analytical decisions across different studies. Moreover, \citet{Wessel2022Quality} found that the outcomes of several RDiT studies are incomparable or report conflicting outcomes. This variability in study outcomes further motivates us to use RDiT to conduct a multiverse analysis, as we believe the different data analytical decisions made might explain the incomparable outcomes. 

\subsection{Primary Study}
\label{sec:primary-study}

We pick one study from the recently conducted RDiT studies in empirical software engineering and use it as a case.
Because of the variation in analytical decisions across RDiT studies, and the conflicting outcomes in these studies~\cite{Wessel2022Quality}, we believe any RDiT study is a good case for a multiverse analysis: \draft{2}{It is a recent data analysis method that has been used to study software engineering. Moreover, \Cref{tab:time-modeling-choices} shows how there is a high degree of researcher freedom in the application of RDiT, and \citet{Wessel2022Quality} shows RDiT studies have conflicting outcomes.} 

The study we select for this multiverse analysis is authored by Cassee~\etal and titled ``\emph{The Silent Helper: The Impact of Continuous Integration on Code Reviews}''~\cite{Cassee:2020}. Selecting this study has a practical advantage, as one of the authors of this study was also involved in the primary study. This familiarity with the primary study and access to an original data archive allowed us to revisit the analytical decisions in more detail. 

For completeness' sake, we briefly summarize the research questions, the data used in the study, the study design, and the study results in the primary study.  

Cassee \etal studies how the adoption of Continuous Integration (CI) by open-source software projects influences the \emph{Communication} within a code review and the number of \emph{Changes} made during a code review. 
Using RDiT, they studied the following four hypotheses: 

\begin{rqs}{}
  H1  & The adoption of CI influences the number of General Comments. \\
  H2  & The adoption of CI influences the number of Review Comments. \\
  H3  & The adoption of CI Influences the number of changes made because of Review Comments. \\
  H4  & The adoption of CI influences the number of commits made during a Code Review. \\
\end{rqs}

H1 and H2 measured the developer's communication, while H3 and H4 measured the changes made during the code-reviewing process.

To verify these hypotheses, Cassee \etal mined code reviews for 685 popular and active GitHub projects that adopted TravisCI, a popular CI service provider~\cite{Rostami:2023}.
To analyze the data, Cassee \etal used an RDiT design. To that end, they created a time series per project centered around the time at which the project started using TravisCI. For each of the four hypotheses, they then fit an RDiT model to understand whether that specific dependent variable was affected by the adoption of TravisCI. \draft{3}{The RDiT model includes the three variables used to model time, and a series of independent variables representing the size and the activity of the community.}

Cassee~\etal finds that a year after the adoption of continuous integration, the average code review is conducted with less communication (\textbf{H1} \& \textbf{H2}). 
Furthermore, no effect of continuous integration on the number of changes made during a code review is found (\textbf{H3} \& \textbf{H4}). 
Therefore, the authors conclude that continuous integration allows developers to perform the same work in a code review while requiring less communication. 

\section{Methodology}
\label{sec:methodology}

In this section, we identify the data analytical decisions of the primary study that we explore in the multiverse analysis (\Cref{sec:multiverse}), and in \Cref{sec:data-analysis}, we describe how we compare the outcomes of the universes. \draft{5}{Additional materials, including the scripts used to generate the multiverse and the outcomes from each universe are available as an online repository on Figshare.}\footnote{\url{https://doi.org/10.6084/m9.figshare.30511331}}

\subsection{Multiverses}
\label{sec:multiverse}

\draft{5}{We follow the steps outlined by \citet{Steegen:2016} to design this multiverse analysis and to analyze the decisions made in the primary study (\citet{Cassee:2020}).}
First, we identify a set of \emph{decision points} in the primary study. For each identified decision point, we then justify reasonable alternatives thereby creating the set of alternative \emph{universes} we explore. \draft{5}{By re-running the experiments of the primary study in these universes, we obtain a set of outcomes we compare to the outcome of the primary study. }

\draft{5}{Designing a meaningful multiverse analysis requires carefully balancing the number of decision points with the alternatives per decision points we consider. \citet{DelGiudice:2021} warns against exploring too many alternative decisions in one multiverse analysis, as combining a large number of arbitrary alternatives creates an exponential number of universes with a large variety of outcomes, complicating the interpretation and usefulness of the results. Therefore, we only select a limited number of decision points from the primary study to explore.}

\draft{5}{
One set of decision points we focus on include the decisions made to model time. We select these decisions because they are already varied in the existing software engineering RDiT literature (cf. \Cref{tab:time-modeling-choices}), showing that in practice, there are many different alternatives being explore across different studies. We combine these decisions with a small number of additional decision points representing the decisions made to aggregate the data and fit the models.   }

\draft{5}{Notably, this excludes other decision points like those related to the processing, filtering, or selection of independent variables. We do not include these decision points for several reasons: some decisions related to data collection cannot be repeated. Re-doing data collection at the time of writing would result in a different dataset, as data on Github will have been altered or removed~\cite{Kalliamvakou:2016}. Secondly, we exclude decisions like the choice of independent variables, as those are decisions for which different theoretical justifications might exist. Thirdly, to ensure interpretability of the multiverse analysis, we exclude any pre-processing decisions not discussed already, to ensure there's no exponential blow-up of explored universes~\cite{DelGiudice:2021}.   }

\subsubsection{Data Processing Decision Points}
\label{sec:data-multiverse}

One group of decision points we explore in this multiverse analysis concerns the modeling of time. We focus on these decisions because they are both crucial to an RDiT study and appear heterogeneous across RDiT studies (See \Cref{sec:rw-rdd}). Additionally, according to \citet{Hausman:2018}, these decisions are related to the sensitivity of RDiT studies. 

In the primary study, each time series consists of 24 30-day periods centered around the point in time when the project adopted continuous integration.
The time series' length (24 time periods) and the resolution (30 days per period) are decisions for which we explore alternative values. 
To account for a period of instability directly before or after the intervention, the primary study excludes 15 days of data before and after the intervention. This is a common decision, seen in several different RDiT studies~\cite{Zhao:2017,Cassee:2020,Wessel:2020}. 
Zhao~\etal justifies this decision based on a manual inspection of some projects. However, as it is unclear how long this instability lasts, and not all RDiT studies use an exclusion period, we also explore several different lengths for the period of instability.

After determining the length and resolution of the time series and the period of instability, the code-reviewing activity is aggregated within a single time period. 
In the original study, a log-scaled mean was used; however, other valid aggregation methods include using a different logarithmic base or taking the median. 
Therefore, the second group of decisions we explore in this multiverse analysis are analytical decisions related to aggregation of data in the time-series.

 \begin{fancylongtblr}[
                    caption = {Parameters and parameter values we analyze.}, 
                    label = {tab:parameters}
                ]{
                    colspec = {X[1]X[3]X[3]}, 
                } 

                Decision & Definition & Values \\ 
                \emph{Time series creation} & Defined as \emph{\#periods} and \emph{periodLength} & $\mathrm{\#periods} \in \{36, 24, 18, 12\}$ and $\mathrm{periodLength}\ \in \{7, 15, 30, 45\}$. \\
                \emph{Period of instability} & Defined as (\emph{daysBefore}, \emph{daysAfter}).  & $\mathrm{daysBefore}, \mathrm{daysAfter} \in \{ (3.5, 3.5), (15, 15), (0, 7), (0, 15) \}$. \\
                \emph{Aggregration} & Defined as \emph{scalingMethod} and \emph{averaging}. & $\mathrm{scalingMethod}\ \in \{\mathrm{original}, \mathrm{ln}, \mathrm{log}_{10} \}$ $\mathrm{averaging}\ \in \{\mathrm{mean}, \mathrm{median}\}$. \\ 
                \emph{Rounding} & Defined as \emph{digitsPrecision}. & $\mathrm{digitsPrecision}\ \in \{\mathrm{unmodified}, 10, 5\}$  

\end{fancylongtblr}

\Cref{tab:parameters} lists all of the decision points related to data processing we explore. \draft{5}{Notably, this excludes other data processing decisions related to, for instance, operationalizations (``\emph{How is the intervention measured?}''). However, we exclude these decisions to reduce the number of studied universes and to ensure the results of the multiverse analysis are meaningful~\cite{DelGiudice:2021}.  }

\subsubsection{Statistical Decision Points}
\label{sec:model-multiverse}

For the decision points related to the statistical models \draft{3}{fit for each hypothesis}, we only focus on a limited set of decision points. There are two factors we focus on: the exclusion threshold for collinearity, which drops collinear \draft{3}{independent} variables \draft{3}{from the RDiT model}, and the fitting algorithm used to fit the model.

 \begin{fancylongtblr}[
                    caption = {Parameters and parameter values we analyze in the statistical model.}, 
                    label = {tab:parameters-model}
                ]{
                    colspec = {X[1]X[3]X[3]}, 
                } 
                Decision & Definition & Values \\
                \draft{1}{\emph{Collinearity}} & \draft{1}{Defined as \emph{vifThreshold}} & \draft{1}{$\mathrm{vifThreshold}\ \in \ \{2.5,\ 5\}$} \\
                \draft{1}{\emph{Fitting algorithm}} & \draft{1}{Defined as \emph{REML}} & \draft{1}{$\mathrm{REML}\ \in\ \{\mathrm{true},\ \mathrm{false}\}$}
\end{fancylongtblr}

\Cref{tab:parameters-model} shows these two decisions and the different values we consider. 

\subsubsection{Instantiating the multiverses}

\draft{5}{In the primary study, a large dataset of code review data originally mined from GitHub was processed into a set of time series, one per project, resulting in a data frame. To create multiverses specified by the decision points in \Cref{tab:parameters,tab:parameters-model}, we modularized the data processing script of the primary study. Each of the decision points described \Cref{sec:data-multiverse} was encoded as an argument of the data processing script. We then ran the modularized script for each unique combination of decision point values, instantiating a set of dataframes representing the multiverses. The modularized data processing script is available in the repository with additional materials.  }

\draft{5}{After instantiating the data frames belonging to each of the decision points, the next step is determining the outcomes in each universe. To do this, we re-run the analysis notebooks from the replication package of the primary study, fitting four models per universe, one model for each hypothesis. By recording the outcomes for each fitted model, we create a mapping from each unique combination of decision points to its corresponding outcome. }

\subsection{Data Analysis}
\label{sec:data-analysis}

\draft{5}{To understand the relation between decision points and outcomes ,we record a limited set of outcomes for each universe. In this Section, we argue why these outcomes are meaningful, how we group outcomes, and how we visualize the relation between decision points and outcomes.} 

\begin{fancylongtblr}
    [
        caption = {The outcomes of the primary study that are recorded in the multiverse analysis.},
        label = {tab:outcomes}
    ]
    {
    colspec = {X[1]X[3]X[3]},
    }

    Outcome & Description & Justification \\
    P-values & The three P-values of the \draft{5}{variables in the model that are used to model the} impact of the intervention. & In the original study, the P-values of these model parameters are used to conclude that the adoption of CI had an effect. \\
    Parameter Sign & Whether the sign of the three model parameters used to model the impact of the intervention are positive or negative    & In the original study, the coefficients of these parameters are used to reason whether trends are increasing or decreasing.  \\
    
\end{fancylongtblr}

For each of those four models, we record the outcomes described in \Cref{tab:outcomes}. \draft{5}{These outcomes capture the information commonly used to interpret RDiT models, and we compare the outcomes in that universe to the outcomes of the primary study. There are additional outcomes we could include in the comparison, like the point estimations of the effect sizes. While we do agree that improvements to  However, this increases the number of possible ways in which outcomes in a universe can differ from the outcomes reported in the primary study. Therefore, we do not include these additional outcomes, and only focus on the outcomes listed \Cref{tab:outcomes}. To further ease the analysis of the outcomes in each universe, to the outcomes of the primary study} we assign the outcome to a \emph{Bucket} representing the similarity of the outcome to the outcome of the primary study. The four groups we use to bucket the universes are: 

\begin{itemize}
    \item \textbf{Full replication}: The significance and sign of the three time-based model parameters are \textbf{all} equal to those of the original study. 
    \item \textbf{Unconfirmed results}: At least one of the claims made in the primary study can not be confirmed in the universe. 
    \item \textbf{Opposite results}: At least one of the time-based variables has significance, while no significance was reported in the primary study. Or the sign of the model parameter is reversed, \ie if the primary study reported a significant increasing trend, a significant decreasing trend is reported in the universe.  
    \item \textbf{Model Fit Failure}: Finally, the decisions made in a universe might result in a dataset, making it impossible to fit a model to the data. While a universe in which models fail to fit usually does not result in a published study, we do want to report these outcomes for the sake of transparency. \draft{3}{While usually, universes in which a model fails to fit would be excluded, we do count these, to understand to what extent empirical software engineering is susceptible to the file drawer bias. }
\end{itemize}

\rf{Not sure about calling them buckets thought, a more formal term might be more specific and apt? We don't want to use outcome though since that is one level down in the original study. Something simple like Analysis Result (AR) with main result and (dependent variable) sub-result (DepVar SAR or AR) for the sub-buckets?}

First, we \draft{5}{assign a universe to an outcome bucket for each of the four dependent variables in the study. Practically, this means that universe $\#x$ might confirm the results for hypothesis 1, but lead to opposite results for hypothesis 2.} 

\draft{5}{As the primary study combines the outcomes of all four hypotheses to conclude that both the communication and the number of changes made in a code review change}, we also assign each universe to a bucket based on the combination of outcomes for each dependent variable. If two or more hypotheses for one universe are in different buckets, we assign the lower bucket of the two. \ie if for one universe the outcome for \emph{General Comments} confirms the primary study, but for  \emph{Commits After Create} the model fails to fit the entire study is placed into the bucket \emph{Model Fit Failure}, as it would not be possible to confirm the findings of the primary study.

\draft{5}{We use several complementary visualizations to explore the relation between universes and outcomes.} Firstly, we count the number of universes belonging to each of the buckets, both for the entire study and for each of the dependent variables. \draft{5}{While this initial counting gives an overview of how many different outcomes there are across the universes, it does not reveal the relation between outcomes and decision points. Therefore, we include the following visualizations:}

\noindent\paragraph{Specification Curves} are a common tool used to analyze multiverse analysis~\cite{Hall:2022}. These specification curves show the relation between individual decisions and the outcome of the multiverse analysis. 

\noindent\paragraph{Stability of Individual Decisions.} For each of the identified decision points, we look at the variability in outcomes, holding all other decisions constant. For instance, for the decision point \emph{periodLength}, we look at the groups of universes where all decisions, except \emph{periodLength}, remain constant. We then count the unique outcomes in each group and visualize the distribution. For decisions that do not affect outcomes, we expect all groups of universes to fall within the same bucket. Whereas, for decisions that affect the outcome, we expect to see large variation in outcomes across groups.

\noindent\paragraph{Timeframe studied.}
When performing an RDiT analysis, \citet{Hausman:2018} recommends against studying a time frame that is too long. As the studied time period becomes longer, it becomes harder to isolate the effect of the intervention on the time series. Therefore, other time-based effects (confounding factors) might introduce more noise, thereby making it more difficult to isolate the impact of the intervention. Therefore, we plot the length of the studied time period in days against the number of different outcomes observed for that timeframe.

\section{Results}

In this multiverse analysis, we explored $3,072$ universes, representing the various reasonable alternatives studied. To understand the relation between data analytical decisions and study outcomes, we fitted $12,288$ RDiT models. In this Section, we discuss the outcomes in each universe and show how particular choices made in constructing the universes influence the outcomes.

\begin{figure}
    \centering
    \includegraphics[width=0.85\linewidth]{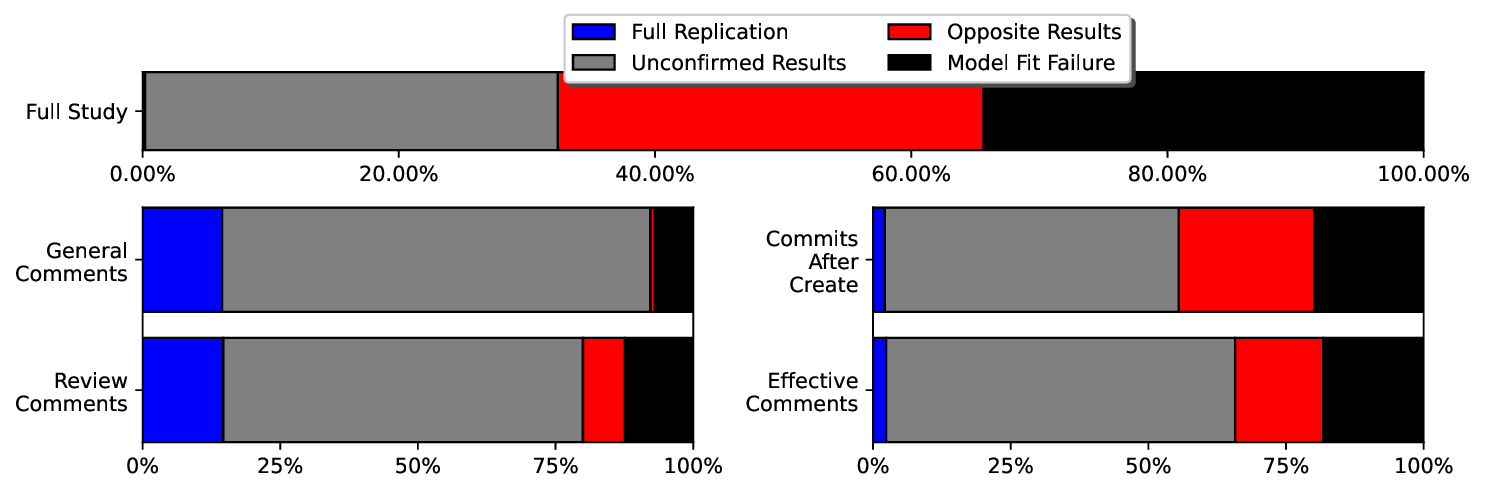}
    \caption{High-level overview of the outcomes for all of the multiverses, per dependent variable. \todo{Highlight this as a red flag?}}
    \label{fig:outcomes-overview}
\end{figure}

\Cref{fig:outcomes-overview} shows a high-level overview of the outcomes. Specifically, it shows how many of the multiverses have an outcome similar to that of the primary study. As discussed in \Cref{sec:data-analysis}, a universe has been assigned to one of the following outcome classes (\emph{Full Replication}, \emph{Unconfirmed Results}, \emph{Opposite Results}, \emph{Model Fit Failure}). 

Most importantly, in only six universes ($0.2\%$), the outcomes for each of the four dependent variables match the outcome of the original study. In $\sim33\%$ of the universes, no significance can be found for at least one of the claims of the original study. More worryingly, in another $\sim33\%$ of the universes, at least one opposite claim can be made (\ie the primary study reports an increasing trend, whereas, in the universe, a decreasing trend can be reported), and finally, in another $\sim33\%$ of the universes at least one of the four models could not be fit. 

Secondly, \Cref{fig:outcomes-overview} shows a large difference in the outcomes for each dependent variable. For the dependent variables related to communication (\emph{General Comments}, \emph{Review Comments}), there are considerably more universes in which the outcome matches that of the primary study ($\sim20\%$). Most importantly, for both General Comments and Review Comments, there is a low number of universes in which opposite results are found. Meanwhile, for the two dependent variables related to the number of changes made during a code review (\emph{Commits After Create}, \emph{Effective Comments}), the number of universes with outcomes similar to the outcome of the primary study is much lower. Even more importantly, for those dependent variables, there are many universes in which claims that are opposite to those of the original study can be made.

Finally, \Cref{fig:outcomes-overview} shows how combining outcomes from multiple models into one conclusion only makes it more likely for alternative decisions to result in different outcomes. As an alternative decision might result in an identical outcome for \emph{General Comments}, but result in a different outcome for \emph{Commits After Create}.

\todo{Model fit failures. We include them, but obviously universes resulting in a model fit failure would not have resulted in the conclusions being made.} \rf{Should we maybe clarify the percentages both when the model fit ones are included and when they are not?}

\implication{Finding}{This multiverse analysis shows a small number of the explored universes lead to an outcome similar to the primary study. This shows how \draft{3}{a high degree of researcher freedom might in and of itself threaten the validity of empirical software engineering studies}. }


\subsection{Specification curve}

\begin{figure}[t]
    \centering
    \includegraphics[width=0.875\textwidth]{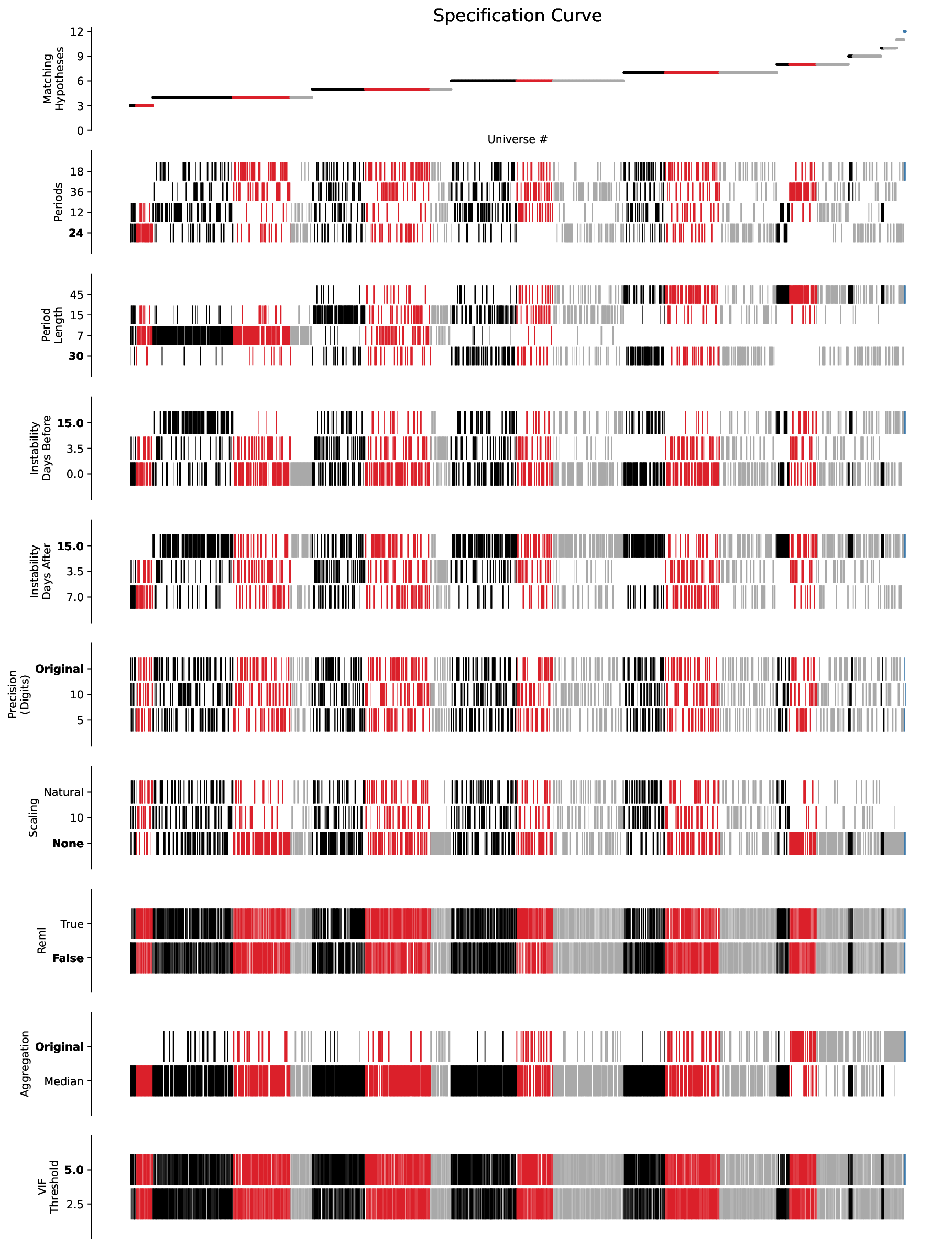}
    \caption{A specification curve showing the number of hypotheses that can be confirmed in each universe, and the relation between a universe, and the decisions made within in the universe.}
    \label{fig:specification-curve-hyp}
\end{figure}

To understand the relation between individual decisions and outcomes, \Cref{fig:specification-curve-hyp} shows a specification curve~\cite{Breznau:2022}. \draft{3}{The plot consists of two parts: The top scatter plot and a row of bar plots. Each point on the top scatterplot represents the number of hypotheses in that universe that match the outcome of the primary study. Each plot below the top plot represents a specific decision, and the presence of a mark on the plot represents the value for the decision used in the universe. The colors represent the outcome categories for the universe, as used in \Cref{fig:outcomes-overview}. }

The distribution of vertical marks for a decision helps understand how a particular choice is related to a specific outcome. 
From the plot, we can, for instance, infer that the value of \emph{Reml} does not appear to be related to any specific outcomes. Because the distribution of marks over the two values for \emph{Reml} is very similar. 
However, for \emph{Period Length} we can see that a value of 7 for never results in a universe where more than 6 of the original 12 hypotheses are replicated. At the same time, universes where a \emph{Period Length} of either 30 or 45 is used appear to confirm more of the findings in the original study, and most importantly, the only universes in which all hypotheses are confirmed are the universes in which \emph{Period Length} is set to 45 days. 

From \Cref{fig:specification-curve-hyp}, we conclude that the study's outcome is strongly influenced by decisions made to model time. When shorter time periods (7 or 15 days) are picked, it becomes impossible to make the same conclusions as the primary study, even when using the same data. \draft{3}{Modeling time is known to be a challenging problem~\cite{Cryer:2008}, and these findings confirm how impactful these decision points can be.}

\draft{3}{More importantly, \Cref{fig:specification-curve-hyp} allows us to conclude that many other data analytical decisions also affect findings. This includes, for instance, the scaling of variables or the method used to aggregate data}. But also the number of datapoints excluded to account for a period of instability. While it is a common practice in RDiT literature to vary these exclusion periods~\cite{Zhao:2017,Kavalar:2019,Wessel:2020, Kinsman:2021,Moharil:2022}, all universes where the exclusion period is shorter than 15 days result in outcomes that are different from those of the primary study. \draft{3}{The specification curve shows how a wide variety of reasonable decisions can make it impossible to confirm the findings of the primary study. }

\implication{Finding}{\draft{5}{The specification curve shows there is a small number of universes with results similar to the primary study, and how decisions related to the modeling of time in RDiT appear to relate to specific outcomes.}}

\subsection{Change Plots}

\begin{figure}[h]
    \centering
    \includegraphics[width=0.82\textwidth]{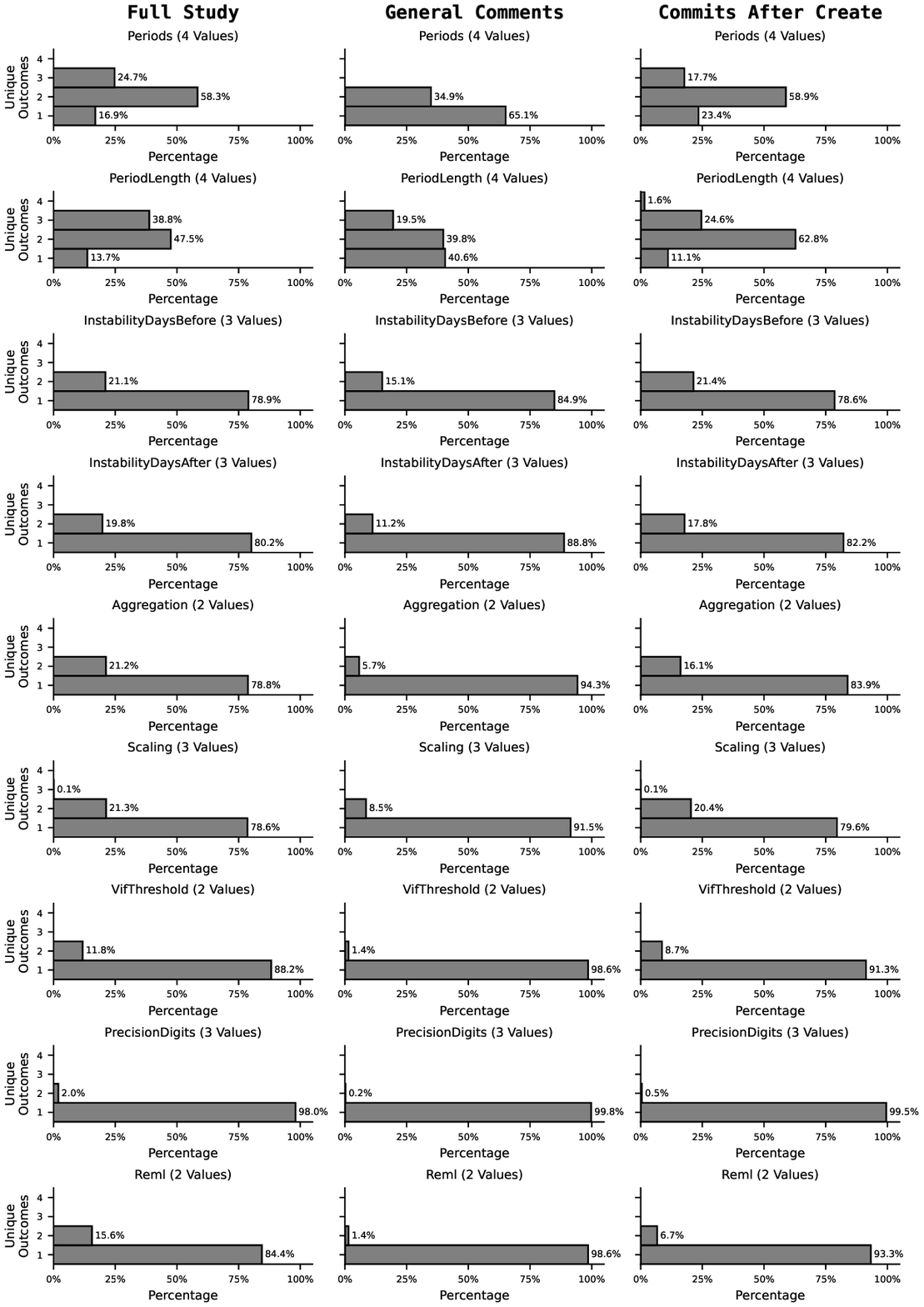}
    \caption{Distribution plots showing the percentage of universes in which changing one decision leads to an alternative outcome.}
    \label{fig:change-plots}
\end{figure}

To complement the specification curve, and to better understand how ``\emph{stable}'' each decision is, we look at all groups of universes where only a single decisions changed. Studying these groups helps quantify how often each specific decision results in an outcome change. 
\Cref{fig:change-plots} plots a set of three histograms for each decision,  each histogram shows the distribution of unique outcomes that can be obtained when changing that decision.  
For stable decisions, we expect to see that a change results in only one outcome being observed, no matter how many alternative values for that decisions are explored. Meanwhile, for unstable decisions the distribution will be skewed towards a higher number of unique outcomes. 

\Cref{fig:change-plots} confirms that changing time-related decisions, \draft{3}{one of the core methodological decisions of RDiT strongly affects outcomes. However, most importantly, it also shows how each decision point can affect the outcome.} Even alternatives for the Vif Threshold, or the Reml value, result in a different outcome in $\sim10\%$ of cases. Finally, most surprisingly, the plot shows that in 2\% of universes, even rounding numbers to a different level of precision can change the outcome. As unlikely as that may seem, we believe it is essential to draw attention to this finding: The digits of precision used to instantiate a data frame are an excellent example of a \textbf{hidden} decision, a choice made by the default setting of a programming environment, and not something you would expect to be influential. Nevertheless, we find that changing it can lead to a different outcome.

\draft{5}{A careful reading of \Cref{fig:change-plots} also shows what the most impactful decisions. For the full study, only changing the \emph{Period Length} resulted in a different outcome in $86.3\%$ of universes. When looking at specific hypotheses (\emph{General Comments} and \emph{Commits After Create}), it remains the most impactful decision. The decision points related to the processing of data, like \emph{Aggregation} and \emph{Scaling} are more stable, but can still lead to meaningfully different outcomes in between $20\%$ to $5\%$ of universes. }

\implication{Finding}{
\Cref{fig:change-plots} shows \draft{3}{there is not a single stable methodological decision}. \draft{5}{In this multiverse analysis \emph{Period Length} is the most impactful decision leading to a different outcome in $86.3\%$ of universes.  \Cref{fig:change-plots} also shows how decisions that are easy to make implicitly, like the number of digits of precision to round to, can also change outcomes.} }

\subsection{Time Curves}

\begin{figure}[t]
    \centering
    \includegraphics[width=0.85\textwidth]{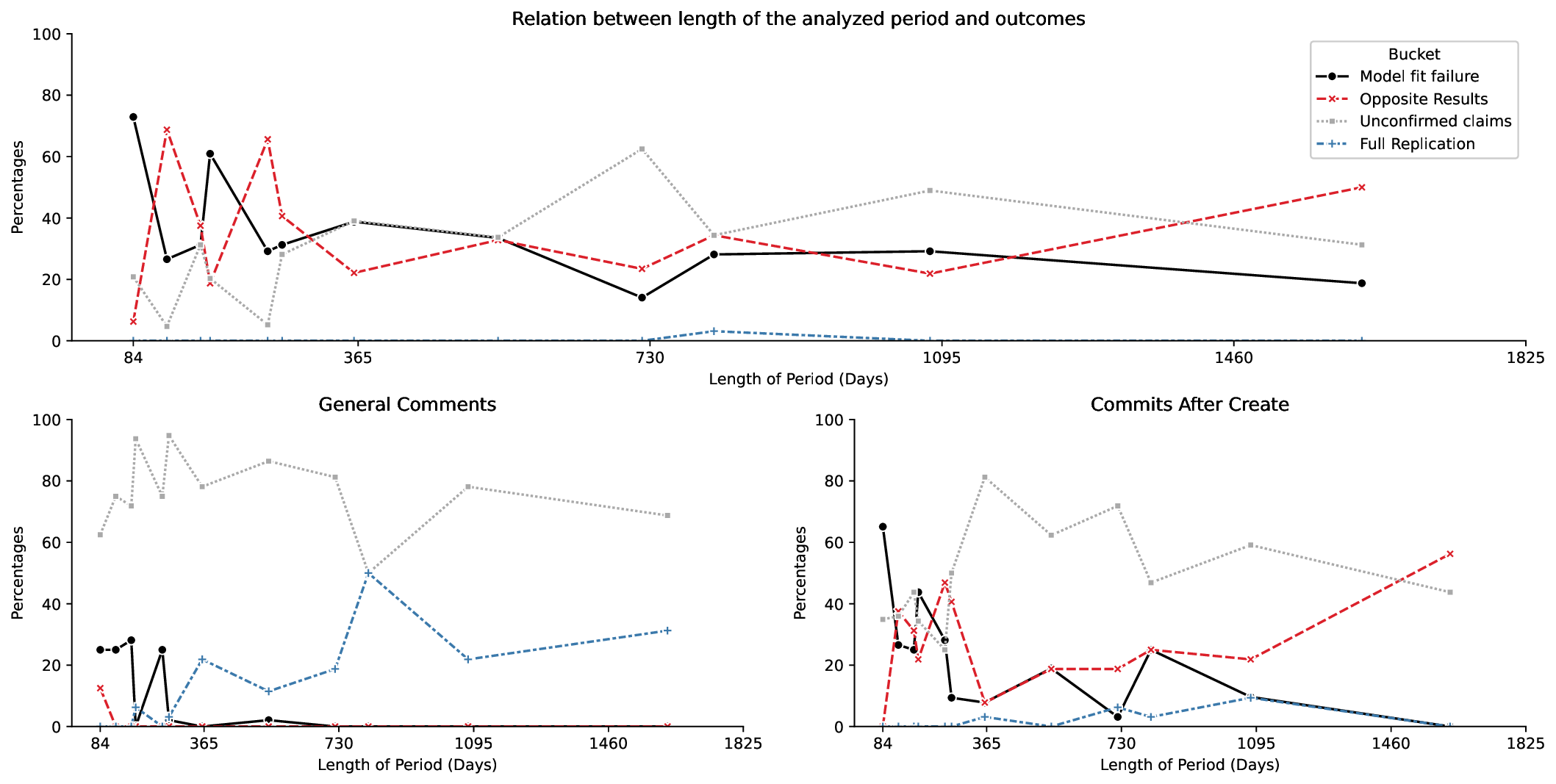}
    \caption{Relation between the length of the studied time period and the outcomes.}
    \label{fig:time-plot}
\end{figure}

\draft{3}{To understand whether a multiverse analysis can help inform how time should be modeled, \Cref{fig:time-plot} plots the relation between the length of the studied time-period, and the outcomes.} The top line plot shows the distribution of outcomes for the full study, while the two bottom plots show outcomes for two dependent variables: \emph{General Comments} and \emph{Commits After Create}.

\Cref{fig:time-plot} shows how outcomes can vary greatly when a shorter period of time is studied -- even resulting in a high number of opposite outcomes.
For periods shorter than a year, the likelihood of obtaining different outcomes is high, as the frequency of different outcomes varies greatly, and even for minor variations in the studied period result in a model failing to fit, or concluding an opposite result. As the length of the total time period increases, the outcomes tend to vary less. However, there is still a noticeable increase in the number of universes in which opposite results can be reported. 

The bottom two plots of \Cref{fig:time-plot} show that the pattern differs for different dependent variables. For \emph{General Comments}, there is a much smaller set of universes that would report opposite results. Meanwhile, this percentage of universes is much higher for \emph{Commits After Create}. This difference across dependent variables and between the entire study and individual dependent variables shows how the same decision might affect different dependent variables. 

This plot shows \draft{3}{the difficulty in picking a stable timeframe} to study using RDiT. The literature recommends studying shorter timeframes---or timeframes that are as short as possible. As \citet{Hausman:2018} argues, the risk of unrelated time-based effects (confounding variables) masking the impact of the dependent variable increases as the studied timeframe becomes longer. However, \draft{3}{In this case, \Cref{fig:time-plot} clearly shows that using a short timeframe introduces a great variation in outcomes. This implies that with the current level of knowledge not enough is known to justify the selection of a single timeframe.}

\section{Discussion}

This study applied a multiverse analysis to a dataset from a published MSR study, mapping nine key methodological decisions and plausible choices for each of them into 3,072 different analysis pipelines. We found that only six of those analytical universes reproduced the findings of the original study. This imbalance reveals a core tension in \draft{3}{empirical software engineering.} While the richness of data is often celebrated as a promise~\cite{Bird:2009, Kalliamvakou:2016}, it also creates analytical fragility by giving researchers a high degree of freedom in decisions. In other words, our findings demonstrate how the greatest promise of \draft{3}{empirical software engineering, and MSR in particular}, the richness of the data, might also be one of its great perils. 
Data analysis pipelines in MSR studies typically require many decisions, often made without explicit guidelines, further amplifying this fragility. Gelman and Loken theorized in their work on the ``garden of forking paths''~\cite{Gelman:2013} \draft{3}{how methodological flexibility might affect robustness. In this paper we confirm this idea, and show how} \textit{a high degree of researcher freedom can cause even a single software engineering dataset and research question to yield many, sometimes conflicting, outcomes}.

Our work extends previous observations of conflicting \draft{4}{empirical software engineering} outcomes (e.g., Wessel et al.’s meta-analysis of RDiT studies~\cite{Wessel2022Quality}) \draft{3}{provides evidence for the fact} that much of that conflict can arise from opaque decision points \draft{3}{that are varied across studies. While we find that the most influential decision points are those related to the modeling of time, we find that \textbf{all} of the studied decision points affect the outcome of the study. In our opinion, the implicit, or hidden, decision points, that are normally not extensively considered are the most problematic. }

Furthermore, we believe that the SE field’s traditional tools for addressing threats to validity are insufficient to address this threat. 
Practices like discussing threats to validity and their mitigation~\cite{feldt2010validity,Wohlin:2012,Menzies:2019,ampatzoglou2019identifying} or discussing study trade-offs~\cite{Robillard:2024} often amount to afterthoughts or are used simply as checklists~\cite{Verdecchia:2023,lago2024threats}. 
Increasing the complexity or type of data analysis~\cite{Furia:2019}, \draft{3}{standardizing the peer review process with empirical standards~\cite{Ralph:2020} or expanding threat discussions will not solve the underlying problem introduced by methodological freedom. Instead, we believe that empirical software engineering research needs more structure to reason about analytical decisions, and especially, to reason about the justification for specific analytical decisions.}

\subsection{Justification Ladder of Analytical Choices}

We propose the ``Justification Ladder of Analytical Choices'' (JLAC), shown in Table~\ref{tab:jlac} to support stronger analytical justifications for data analysis choices. This model, developed from our experience with the multiverse analysis presented in this paper, offers empirical software engineering researchers a structured way to classify and, ideally, strengthen justifications for methodological decisions. \draft{3}{This justification ladder complements existing initiatives to improve methodological rigor. For instance, where the empirical standard focuses on \textbf{what} should be justified~\cite{Ralph:2020}, the justification ladder helps distinguish specific \textbf{types} of justification.} Below, we describe the JLAC in more detail and explain how it can help researchers make their analytical decisions more transparent and build greater trust in MSR and, more generally, in empirical software engineering studies. 

\begin{fancylongtblr}
    [
        caption = {The Justification Ladder of Analytical Choices (JLAC) with stronger and more preferable levels of justification on higher levels (5) of the ladder  .\todo{Maybe the table needs some examples?} \rf{I suggest we put them in a table of their own and try to make them specific to one of our variables around time}},
        label = {tab:jlac}
    ]
    {
    colspec = {r X[1]X[5]},
    }

    Level & Name & Description \\

    5 & Causal & These decisions are made based on causal models or theories of the problem, or domain, under study.  \\
    4 & Empirical & These decisions are justified based on properties of the data under study, but these decisions are not necessarily linked to a causal model \\
    3 & Heuristic & These are decisions made based on general heuristics, often made based on reasoning that it is not directly tied to the domain, or the research question being studied. \\
    2 & Conventional & These decisions are made based on existing conventions, where the justification of the decision is made because it is common practice in the field. \\
    1 & Implicit & These are decisions that are made implicitly, without the researcher being aware the decisions is being made. These could be defaults of the statistics package being used. \\
    
\end{fancylongtblr}

Following the scientific method, researchers begin with a theory to generate hypotheses, then design studies and analyses to test those hypotheses—each step producing falsifiable predictions. When analytic choices ((e.g., variable selection, data transformation, modeling relationships) are derived from a detailed causal theory, two key benefits arise: (a) assumptions are explicitly articulated, tying every transformation or filter to a hypothesized mechanism rather than leaving them to software defaults or chance; and (b) reproducibility is enhanced, as other teams applying the same causal model are more likely to make similar decisions.

In contrast, selecting analyses based on (potentially) empirical quirks of a dataset risks circular reasoning and overfitting. By grounding analytic pipelines in causal models, we transform them into disciplined, theory-driven experiments. This is why detailed scientific theories should be the preferred basis for analytic choices.

Decisions made implicitly or without explicit discussion are more likely to introduce unintended biases. Similarly, choices based on conventions or general heuristics often drift away from the specific domain or context of a study, providing weaker justifications than the data at hand. However, while causal and empirical justifications are preferable, we recognize that theoretical knowledge may not always be available or practical to apply. Thus, our model organizes types of justifications into a hierarchy, from weaker to stronger, reflecting their desirability.

\Cref{tab:jlac} lists the types of justifications we distinguish. We argue that empirical software engineering researchers should aim to ground data analytic decisions as high up this justification ladder as possible, ideally using verified causal models to inform analytical choices. 

Prior work supports this need to ``move up the ladder''.
Kale~\etal found that many published studies base decisions on conventions or leave them undiscussed~\cite{Kale:2019}. Conflicting operationalizations~\cite{Wyrich:2024} and divergent methodological decisions for similar research questions complicate meta-studies, \draft{3}{and \citet{Wyrich:2024:pixelated-threats} emphasize the need to make data-driven decisions to quantify threats to validity.} This further highlights the need for better tools to help researchers justify their choices more rigorously, thereby enhancing the reliability of individual studies and enabling meta analyses to more effectively compare and synthesize findings.

Furthermore, \citet{Gelman:2013} argues that when analytic decisions are shaped by the data itself, findings should be treated as exploratory rather than confirmatory. Our multiverse analysis suggests that this caution also applies to MSR studies that rely on implicit defaults or widely adopted conventions without explicit and strong justification. In such cases, even studies presented as confirmatory may simply reflect one path through a forest of equally plausible analyses. Following \citet{Gelman:2013}, we contend that these studies might be better characterized as exploratory. We therefore recommend that authors acknowledge these dependencies and, where possible, reinforce their confirmatory claims with sensitivity analyses or clear, theory-driven rationales for their analytic choices.

\subsection{Applying JLAC to RDiT Studies}
\label{subsec:rdits-motivations}

\draft{5}{
The Justification Ladder of Analytical Choices (JLAC) can be concretely illustrated through the motivations typically reported for the time-series design decisions in Regression Discontinuity in Time (RDiT) studies. 
In Table~\ref{tab:jlac-timeseries} we map the three temporal decisions---number of periods, period length, and instability exclusion window---to their corresponding levels on the JLAC.\footnote{\draft{6}{It is important to note that our mapping is based on the interpretations of each of the manuscripts; it might not take into account justifications not explicitly listed in the manuscript.}}
By examining how prior RDiT studies in software engineering justify these choices, we can assess prevailing justification practices and identify opportunities to strengthen them.} 

\paragraph{Number of periods.}
\draft{5}{
Some RDiT studies provide at least a partial rationale for their chosen number of periods, often linked to data availability or pragmatic considerations. 
For instance, Cassee~\cite{Cassee:2020} modeled twelve months before and after the introduction of continuous integration (CI) to ensure that each bucket contained a sufficient number of projects for model estimation (\textit{Level~3, Heuristic}).
Zimmermann~\cite{Zimmermann:2019} adopted a conservative specification using a smaller bandwidth to minimize bias but tested an extended window (511 days on each side) as a robustness check (\textit{Level~4, Empirical}). 
Walden~\cite{Walden:2020} selected 25~months on each side to balance sufficient data points against the risk of confounding factors far from the cutoff (\textit{Level~3, Heuristic}). 
Saraiva~\cite{Saraiva:2023} used 24~versions (12 before and 12 after the split event) to maintain symmetry around the intervention point (\textit{Level~3, Heuristic}). 
Other RDiT papers provide no motivation for this decision. }

\paragraph{Period length.}
\draft{5}{
None of the reviewed RDiT papers explicitly justify their choice of period length. In all examined cases, the period (e.g., 30~days) seems to be inherited from established practice or software defaults, with no discussion of alternative temporal granularities or their possible impact on the results. According to the JLAC, this lack of explicit reasoning corresponds to Level~1 (Implicit). One might argue that the near-universal use of similar period lengths makes this a de facto community convention and thus closer to Level~2. However, when the choice is neither mentioned nor reflected upon, it is difficult to claim that it was consciously made, so we consider Level~1 the more appropriate classification.}

\paragraph{Instability exclusion window.}
\draft{5}{
The choice of how much data to exclude around the intervention event \draft{6}{is sometimes justified empirically, but it is mostly based on anecdotal observations, partial observations, or on convention. A prime example of empirical validation is the donut RDDs investigated by \citet{Zimmermann:2019}.} Meanwhile, Zhao~\cite{Zhao:2017} excluded one month of data centered on the CI adoption event, motivated by informal observations of restructuring activity during this period (e.g., changes in build systems or dependencies; Level~3, heuristic). Although one might argue this approaches Level~4, there is no systematic, quantitative analysis to support that classification, so we retain it as Level~3. Cassee~\cite{Cassee:2020}, Kinsman~\cite{Kinsman:2021}, and Saraiva~\cite{Saraiva:2023} then followed this convention, excluding 15~days before and after the intervention (Level~2, Conventional). The other studies did not motivate or discuss this choice at all.
}

\begin{table}[h]
\centering
\caption{Application of the Justification Ladder of Analytical Choices (JLAC) to the main time-series design decisions in RDiT studies.}
\label{tab:jlac-timeseries}
\begin{tabular}{p{3cm}p{4cm}p{3cm}p{3cm}}
\toprule
\textbf{Decision Point} & \textbf{Example Choices Motivations} & \textbf{JLAC Levels} & \textbf{Common Limitations} \\
\midrule
Number of periods & Ensuring sufficient data or balancing sample size and bias \cite{Cassee:2020} & 
\textbf{4}: \cite{Zimmermann:2019}; \textbf{3}: \cite{Cassee:2020,Walden:2020,Saraiva:2023}; \textbf{1}: \cite{Zhao:2017,Guo:2019,Kavalar:2019,Wessel:2020,Kinsman:2021,Trockman:2018,Moldon:2021,Moharil:2022,Li:2023,Ayoup:2022} & Often not discussed (i.e. level 1, implicit), driven by heuristic reasoning on ``balancing'' data or the number of periods (3), or, in one case, based on simulation on data (4). \\[3pt]
Period length & Typically 30 days but rarely discussed nor motivated & \textbf{1}: \cite{Zhao:2017,Guo:2019,Kavalar:2019,Wessel:2020,Cassee:2020,Walden:2020,Zimmermann:2019,Kinsman:2021,Trockman:2018,Moldon:2021,Moharil:2022,Li:2023,Ayoup:2022,Saraiva:2023} & No motivation given, so likely chosen for convenience\slash convention, ignoring alternative granularities or behavioral rhythms \\[3pt]
Instability exclusion window & Exclusion of 15 days around the intervention to remove transitional noise, \draft{6}{based on empirical testing for \citet{Zimmermann:2019}}, and anectdotal observations in some projects~\cite{Zhao:2017}, which then became convention~\cite{Cassee:2020,Kinsman:2021,Saraiva:2023} & \draft{6}{\textbf{4}: \cite{Zimmermann:2019}}; \textbf{3}: \cite{Zhao:2017}; \textbf{2}: \cite{Cassee:2020,Kinsman:2021,Saraiva:2023} \textbf{1}: \cite{Guo:2019,Kavalar:2019,Walden:2020,Wessel:2020,Trockman:2018,Moldon:2021,Moharil:2022,Li:2023,Ayoup:2022,Saraiva:2023} & \draft{6}{Mostly motivated} by anecdotal observations rather than quantitative or theoretical grounding, rest motivated by convention or didn't motivate. \\
\bottomrule
\end{tabular}
\end{table}

\draft{5}{Beyond the temporal design choices, five additional analytical decisions were explored in the multiverse analysis: aggregation method, scaling transformation, rounding precision, variance-inflation threshold, model-fitting algorithm (REML or not). 
While typically not specifically discussed or motivated in the RDiT literature, these decisions largely reflect standard statistical or implementation defaults. 
For example, the aggregation and scaling choices are often heuristic (Level~3) or by convention (Level~2), while rounding and estimation method defaults remain implicit (Level~1). 
Together, these patterns reinforce that analytical justification in current software-engineering RDiT studies is concentrated in the lower and middle levels of the JLAC (Levels~1–3), indicating opportunity for methodological strengthening.}

\draft{5}{Because our JLAC assessment is based solely on what is reported in published manuscripts, it inevitably offers an incomplete picture of how methodological decisions were actually made. Some justifications may have been omitted due to page limits, stylistic choices, or assumptions about what readers consider obvious. For this reason, we do not position the JLAC as a tool for retrospectively judging existing studies; instead, we see it as a forward-looking aid for researchers to structure and scrutinize their own decision points in ongoing and future work. When planning and conducting quantitative analyses, we hope the JLAC will help researchers articulate, compare, and strengthen their methodological justifications in a more systematic way.}

\subsection{Strengthening Justifications of Analytical Decisions}

Our findings and the argument above raise an important question: \emph{What can be done to strengthen the justification of analytical decisions?}

We start by discussing several ways to strengthen methodological decisions in RDiT studies.
For example, the \emph{number of periods} could be empirically validated against the minimum data window needed for stable coefficient estimation; \emph{period length} could be aligned with theorized or empirically investigated release or contribution cycles in software development; and the \emph{instability exclusion windows} could be derived from quantitative analyses of recovery or adaptation time after interventions. 
Grounding such decisions in either observed data patterns or process theory would elevate their justification from Levels~1–3 toward Levels~4 and potentially even 5, enhancing both methodological transparency and the interpretability of RDiT results.

\draft{5}{Beyond RDiT, several tools and methods can help move justifications for data-analytical decisions higher up the justification ladder. We envision this as a three-step process. First, identify key decision points. Second, make these decision points and their viable alternatives explicit. Third, evaluate and compare the alternatives to select those most appropriate for the study. Researchers can rely on a range of strategies at this third step, from drawing on established theories to applying quantitative techniques such as simulation-based evaluation.}

One way to inform and justify methodological decisions is to draw on existing scientific theory~\cite{Stol:2015}. Well-developed theories provide a strong foundation for choosing models, variables, and analytical strategies. Practical approaches for building and evaluating such theories include causal modeling~\cite{McElreath:2018, Furia:2019, Furia:2023} and structural equation modeling~\cite{Russo:2023, Trinkenreich:2023}, both of which have already been successfully applied in software engineering contexts.

Although justifying every methodological decision with established theory would be ideal, it is rarely feasible in practice. Fortunately, several other approaches can substantially strengthen justifications and guide decisions. One promising option is to use simulated data to test whether a proposed analysis method is appropriate~\citet{Hartel:2022, Hartel:2023}; by generating data with known effect sizes, researchers can compare alternative choices at key decision points—for example, in an RDiT study, confirming that a specific combination of time-modeling decisions can recover a small effect in noisy, multi-level data. A second approach is pre-registration~\cite{ernst2023registered}, which reduces the likelihood that study outcomes depend on post hoc analytical choices. A third is mixed-methods research, where MSR studies are complemented with other methods to triangulate and confirm findings~\cite{Storey:2025}.

Finally, our study highlights the value of multiverse analyses--systematically exploring alternative analytic decisions--as an effective ``smoke test'' for assessing the sensitivity of study outcomes. In domains with high methodological freedom, multiverse analyses can map out the range of plausible conclusions that can be drawn from the same data. If a multiverse analysis shows that a single data source supports conflicting conclusions under reasonable alternatives, this should be interpreted as a \textbf{red flag}, signalling that domain knowledge is too limited to support the planned analysis with confidence. In such cases, the research community should aim to reduce unnecessary methodological freedom, constrain analytical choices, and move higher up the ladder of justifications by applying the tools and methods discussed in this section.

\section{Limitations}

The multiverse analysis described in this manuscript has its own set of limitations, which stem from our methodological decisions rather than the limitations of the primary study itself.

First, our analysis is restricted by issues of \textbf{ external validity}. By focusing on a single MSR study, we cannot assume that our findings generalize across the entire MSR literature. We mitigated this risk by selecting a research method, trace-data analysis, that is widely used in software engineering, and by building on a meta study that has already documented conflicting outcomes across RDiT applications~\cite{Wessel2022Quality}. Nonetheless, our conclusions remain specific to this case.

Second, we were unable to revisit all data collection choices made in the original study, particularly the criteria for selecting repositories (e.g., minimum stars, contributor counts). This limitation also touches on \textbf{external validity}: although our multiverse explores many decision paths, it does not encompass the full space of repository-selection strategies used across MSR. However, the dependencies we uncover between analytic decisions and outcomes hold for the data set at hand.

Third, a key strength of a multiverse analysis is that it can vary operational definitions, but this very variation introduces a risk to \textbf{construct validity}. By systematically changing measures of a specific construct, thresholds for inclusion, and data-cleaning rules, some analytic branches may no longer align well with the original theoretical constructs. Future work could empirically assess which combinations of operational choices best map onto the underlying constructs of interest.

Finally, and clearly, our multiverse analysis does not address the limitations of the primary study’s \textbf{internal validity} (e.g., causal claims, confounding factors) and therefore does not resolve any threats to that study's design.

Taken together, these limitations underscore that while a multiverse analysis offers better transparency into the effect of methodological choices on results, it also inherits, and in some cases may amplify, the validity concerns of both the primary study and the analytic processes employed.

\section{Conclusion}

The replication crisis, triggered by a series of high-profile failed replications, has become a cornerstone of meta-scientific research on how to improve the reliability of science. \citet{Gelman:2013} proposed that one contributor to this crisis is the wide methodological freedom researchers often have when making analytic decisions. 
\citet{Silberzahn:2018} empirically confirmed this by demonstrating that different research teams, given the same dataset and research question, frequently reach different conclusions because of variations in their methodological choices.

\draft{3}{Empirical software engineering, and Mining Software Repositories (MSR) research in particular,} shares this freedom. The richness of data is often presented as a great promises~\cite{Bird:2009,Kalliamvakou:2016}, but it also raises concerns about how methodological decisions might shape study outcomes. To investigate this, we selected a published MSR study that employed a data-analytic method allowing considerable researcher freedom. We then conducted a multiverse analysis: systematically exploring reasonable alternatives to the original study’s methodological decisions and reporting the outcomes across these alternatives.  

Our analysis considered 3,072 alternative ``universes,'' each corresponding to a unique combination of nine methodological decisions. We found that every one of the nine decisions could change the study's results. In fact, only 6 out of the 3,072 explored universes replicated the original study’s findings. This highlights that choices in processing and \draft{3}{analyzing post-hoc observational data commonly used in empirical software engineering} data may influence outcomes far more than previously assumed.

These results indicate a need for stronger methodological rigor in \draft{3}{empirical software engineering}. Analytical choices should be grounded in causal or theoretical frameworks when possible. When such grounding proves infeasible, validation via simulation, mixed-methods research with triangulation, or targeted sensitivity checks can help avoid fragile conclusions. Authors might consider conducting and reporting multiverse analyses to reveal how different analytical paths affect their findings. They can also adopt the Justification Ladder of Analytical Choices, which we proposed based on our experiences conducting this multiverse analysis, to document and defend their analytical decisions. Finally, specifically for RDiT studies (and other time-dependent analyses in MSR), defining and justifying domain-specific guidelines for period length and instability exclusion can reduce outcome volatility. Since we have presented only a single case future work should consider multiverse analysis also of other MSR studies and for other types of design.

\section*{Acknowledgments}

We acknowledge support from the Science Council (Vetenskapsrådet, contract id 2020-05272) for the project "Automated Boundary Testing for AI/ML models" and from WASP for the project "Bound-Miner". We also want to sincerely thank Julian Frattini, Alexander Serebrenik, and Richard Torkar for their feedback on an earlier version of this paper! 

\bibliographystyle{ACM-Reference-Format}
\bibliography{references}

\end{document}